\documentclass[twocolumn, trackchanges]{aastex701}
\usepackage{amsmath}
\usepackage{newtxtext,newtxmath}
\usepackage{xcolor}
\usepackage{adjustbox}
\usepackage{graphicx}
\usepackage{amsmath}
\usepackage{footnote}
\usepackage{amssymb}
\usepackage{multirow}
\usepackage{threeparttable}

\def \rxte{{\it RXTE}}

\def \xmm {{\it XMM}-Newton}
\def \src{{4U~1323-62}}
\def \obj{{4U~1323-62}}

\def \nustar{{\it NuSTAR}}

\def \nicer{{\it NICER}}

\def \astrosat{\textit{AstroSat}}

\shorttitle{Burst oscillation in \src{}}
\shortauthors{Mandal \& Naik}

\begin{document}

\title{Detection of possible burst oscillation in the neutron star low-mass X-ray binary \src}

\correspondingauthor{Manoj Mandal}
\email{manojmandal213@gmail.com}
\author[orcid=0000-0002-1894-9084]{Manoj Mandal}
\affiliation{Astronomy and Astrophysics Division, Physical Research Laboratory, Navrangpura, Ahmedabad - 380009, Gujarat, India}
\email{manojmandal213@gmail.com}

\author[orcid=0000-0003-2865-4666]{Sachindra Naik} 
\affiliation{Astronomy and Astrophysics Division, Physical Research Laboratory, Navrangpura, Ahmedabad - 380009, Gujarat, India}
\email{snaik@prl.res.in}

\begin{abstract}
Burst oscillations observed during thermonuclear X-ray bursts arise from asymmetric brightness patterns on the neutron star surface and provide a direct probe of the neutron star spin frequency. We present a detailed timing analysis of the neutron star low-mass X-ray binary \src~ using 2024 observations with \xmm~ and \nustar~ observatories. We identify nine thermonuclear X-ray bursts in the \xmm/EPIC-pn data, along with eclipsing dips in the light curve. One of the \xmm~ bursts exhibits a rare doublet structure. In addition, \nustar~ detects six bursts, four of which occur simultaneously with those observed with \xmm. We identify a possible burst oscillation signal at $\sim$611.5 Hz in the \xmm\ data.  The strongest oscillation, detected during the primary burst of the doublet burst, reaches a maximum $Z_1^{2}$ power of $\sim35$. An analytical estimate accounting for the searched frequency and time intervals gives a significance of $\sim3.0\sigma$, whereas independent Monte Carlo simulations yield a more robust global significance of only $\sim2.4\sigma$. We therefore interpret the signal as a tentative detection of burst oscillation. The folded pulse profile in the 0.5--10 keV band is well described by a sinusoid, with a fractional rms amplitude of $\sim30\pm6$\%. The oscillation frequency corresponds to a neutron-star spin period of $\sim$1.635 ms, suggesting that \src\ may harbor a rapidly rotating millisecond neutron star.
\end{abstract}
\section{Introduction}
\label{intro}
A neutron star low-mass X-ray binary (LMXB) consists of a neutron star (NS) accreting matter from a low-mass companion star (typically $\lesssim 1,M_{\odot}$) in a compact binary system. In these systems, matter from the companion star is transferred through Roche-lobe overflow and accreted onto the neutron star through an accretion disc. In the case of weakly magnetized (B$\simeq$10$^{7-9}$ G; \citealt{Ca09, Mu15}) NS LMXB systems, the accretion disc can extend close to the neutron star surface, allowing the accreted material to spread over the surface rather than being funneled onto the magnetic poles. As the accreted matter accumulates on the surface of the neutron star, the temperature and pressure at the base of the accreted layer gradually increase. Once critical conditions are reached, unstable thermonuclear burning of hydrogen and/or helium is triggered, producing sudden intense X-ray flashes known as thermonuclear (Type~I) X-ray bursts \citep{Le93, Ga06}. These bursts typically last for several tens of seconds and release energies of the order of $\sim10^{39}$ erg in a single event. They are characterized by a rapid rise lasting a few seconds, followed by a slower exponential decay. During the bursts, lighter elements are burned into heavier nuclei through a chain of thermonuclear reactions \citep{Le93, St03, Ga21}.

In addition to the spectral and temporal evolution, thermonuclear bursts often exhibit nearly coherent brightness modulations known as burst oscillations. Burst oscillations have been detected in several NS LMXBs during thermonuclear X-ray bursts, typically at frequencies in the range of 200--600 Hz, consistent with the spin frequency of the neutron stars \citep{Watts2012, Ga08}. The detection of burst oscillations provides a useful method for measuring the spin frequency of neutron stars. These oscillations are generally attributed to asymmetric emission patterns on the neutron star surface arising from unstable nuclear burning. They also provide valuable insight into the physics of thermonuclear flame spreading and the behavior of ultra-dense matter under extreme physical conditions \citep{Strohmayer2006, Watts2012}. However, the physical origin of burst oscillations remains uncertain, and only about 10\% of thermonuclear bursts exhibit such signals in high time-resolution observations \citep{Ga08, Watts2012}. For a comprehensive review of burst oscillation sources and their properties, see \citet{Watts2012}.

In some thermonuclear bursts, the luminosity reaches the Eddington limit, causing the neutron star photosphere to expand temporarily above the stellar surface. Such events are known as photospheric radius expansion (PRE) bursts \citep{Le93, Ta84, Ga21}. During the expansion phase, the photospheric radius increases while the blackbody temperature decreases, with the bolometric flux remaining nearly constant at the Eddington luminosity. As the photosphere subsequently contracts back toward the neutron star surface, the temperature rises again; this is known as the touchdown phase. The PRE bursts are particularly important because they provide a means to constrain the neutron star radius \citep{Guver2010}. These measurements place important constraints on the equation of state (EoS) of ultra-dense matter and improve our understanding of the internal composition and structure of neutron stars.

The blackbody model \citep{Va78, Ku03}, which assumes ideal thermal emission from the neutron star surface, is commonly used to describe burst spectra \citep{Mandal2023, Malacaria2025}. However, significant soft and hard excess emission is often observed during intense bursts \citep{Worpel2013}, indicating deviations from a pure blackbody emission. This excess may result from enhanced persistent emission driven by Poynting-Robertson drag \citep{Walker1992, Worpel2013}, and/or from reprocessing of burst photons in the accretion disc \citep{Jaisawal2024, Yu24, 4U1702, Mandal2025, Mandal2026MNRAS}. Indeed, several intense bursts exhibit emission and absorption features, providing strong evidence for reflection from a photoionized accretion disc \citep{Ke17, De18, Ba04}. 

The neutron star \src~ (also known as XB~1323-619) was discovered in 1970 by {\it Uhuru} and {\it Ariel~V} missions \citep{Forman1978, Warwick1981}. It was classified as a neutron star based on observations and findings from {\it EXOSAT} \citep{vanderklis1985}. The neutron star LMXB \src~ exhibits type-I X-ray bursts and periodic dipping behavior, as previously observed with EXOSAT \citep{vanderklis1985}. These dips are likely caused by periodic obscuration of the central X-ray source by structures in the accretion disk and the companion star \citep{White1982}. The inclination angle is reported to be high (60-80$^\circ$) \citep{Frank1987}. Earlier, \rxte~ observations revealed multiple X-ray bursts and nearly periodic dips \citep{Barnard2001}. The source also shows $\sim$1~Hz quasi-periodic oscillations (QPOs) during persistent and burst emission \citep{Jonker2000}. Additionally, a distinct $\sim$1~Hz QPO has been reported during the dipping phases, possibly associated with quasi-periodic obscuration at high binary inclination, consistent with the characteristics of dipping sources such as \src. The results from \astrosat~ and \nustar~ observations also showed multiple thermonuclear bursts \citep{Bhulla2020, Bhattacharya2026}. The low-frequency QPO at 1 Hz is also detected in \astrosat~ observations. A recent work on \xmm~ and \nustar~ campaign confirmed the source's dipping activity and orbital period. The authors also reported a long-term $\sim$10 yr modulation in the burst rate and detected a persistent $\sim$0.9 Hz QPO outside bursts and dips. \citep{Boztepe2026}. They also performed a detailed time-resolved burst spectroscopy to investigate burst properties and possible coronal cooling induced by burst irradiation.

 During the \rxte~ era, \citet{Bilous2019} carried out a comprehensive timing and spectral analysis of thermonuclear X-ray bursts from several neutron star LMXBs, including \src. Owing to its infrequent X-ray activity, \src~ has been studied less frequently compared to many other NS LMXBs This motivates a detailed broadband investigation of the source using simultaneous observations from \xmm~ and \nustar. Thermonuclear X-ray bursts from this source have previously been studied using \rxte{} \citep{Bilous2019}, {\it EXOSAT} \citep{vanderklis1985}, \astrosat{} \citep{Bhulla2020}, and \nustar{} \citep{Bhattacharya2026}. However, detailed studies of burst properties with sensitive soft X-ray coverage below $\sim$3 keV remain limited, even though this energy range provides important insight into the burst and accretion physics. Since burst emission predominantly peaks in the soft X-ray band (below $\sim$6 keV), observations with instruments having strong low-energy sensitivity are particularly valuable.

In the present work, we report the detection of thermonuclear bursts from \src~ using \xmm~ and \nustar~ observations. We perform a comprehensive timing analysis of these bursts, including a search for burst oscillation signatures in individual bursts observed with both missions. The paper is organized as follows. Observations and data reduction procedures are described in Section~\ref{obs}. Section~\ref{res} presents the results of the spectral and timing analyses, while the implications of these findings are discussed in Section~\ref{dis}.
\begin{table}
\centering
\caption{Summary of observations and X-ray bursts detected from \obj{} using \xmm, and \nustar{} observations. A total of 15 bursts are observed with \xmm~ and \nustar, and 4 bursts are simultaneously detected in \xmm~ and \nustar.}
\begin{tabular}{llccc} 
\hline	
Observatory	& Date of       & Observation & Exposure & No. of     \\
		      &observation    & ID           & (ks)    & bursts     \\
\hline
{\it NuSTAR}    &2024-08-07    & 31001021002	    &90     &6  \\
{\xmm}          &2024-08-07    & 0935800201	    &135     &9   \\
\hline
\label{tab:log_table_burst}
	\end{tabular}
\end{table}
 \begin{figure*}
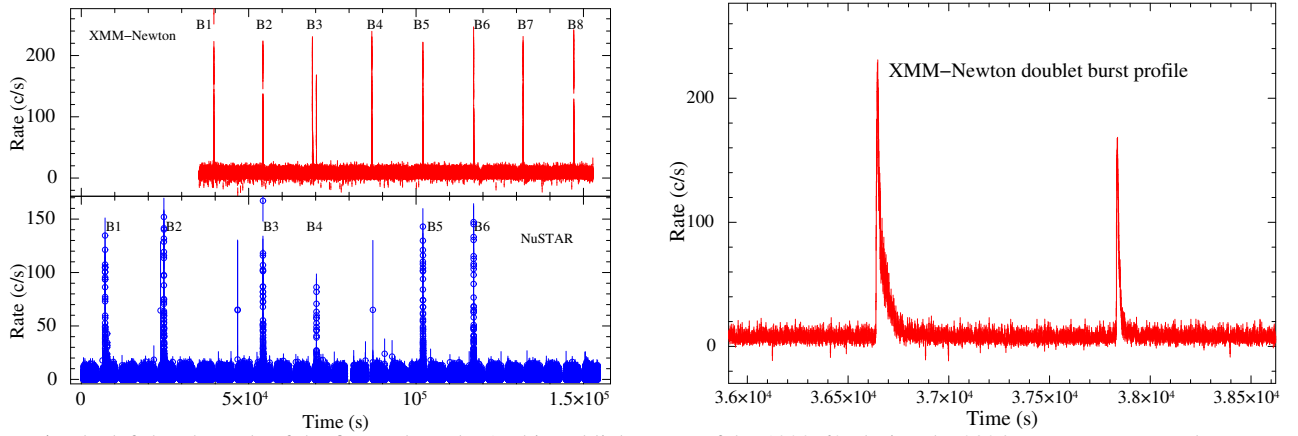

\centering{
\includegraphics[width=0.35\textwidth, angle=270]{lc_simultaneous_4U.eps}
\includegraphics[width=0.35\textwidth, angle=270]{doublet_profile.eps}
\caption{The left-hand panels of the figure show the 1-s binned light curve of \src~ during the 2024 \xmm~ and \nustar~ observations. During these observations, a total of 15 thermonuclear X-ray bursts are detected, and 4 bursts are simultaneously detected with both instruments. Each burst is assigned an identifier. One of the bursts observed with \xmm~ (marked as B3) exhibits a doublet burst profile, as shown in the right-hand panel.}
\label{fig:burst_combined}}
\end{figure*}
\section{Observation and data analysis}
\label{obs}
In this work, we used publicly available data from observations of \obj{} with the  \xmm{}, and \nustar{} observatories. The \nustar~ data are reduced using {\tt HEASoft} version 6.36 along with the latest versions of caldb. 
\subsection{\xmm{} observation}
The European Photon Imaging Camera (EPIC; 0.1–15 keV) is mounted at the focal planes of the three $\sim$1500 cm$^2$ X-ray telescopes of \xmm{} \citep{Ja01}. \obj{} was observed on 7 August 2024 for an exposure of $\sim$135 ks (Obs. ID-0935800201, PI: Tolga Guver). The details of the observation are summarized in Table~\ref{tab:log_table_burst}. The observation was performed in EPIC-PN timing mode. The \xmm~ data are processed using the Science Analysis System ({\tt SAS}) v20.0.0. EPIC-PN event files are generated with {\tt epproc}, and the events are filtered using {\tt evselect} with FLAG = 0 and PATTERN $\leq 4$. Source and background light curves and spectra are extracted using {\tt evselect}. The source events are extracted from a 12-pixel-wide strip centered on the source position RAWX = 36 (i.e., RAWX in [30:42]), and the background region is selected from a source-free region in RAWX columns [2:15] from the same CCD. The EPIC-pn timing-mode data are barycenter-corrected using {\tt barycen}, adopting the JPL DE405 solar system ephemeris and source coordinates RA (J2000) = 201.6543$^{\circ}$ and Dec (J2000) = $-62.1358^{\circ}$ \citep{Gambino2016}.
\subsection{\nustar{} observation}
The Nuclear Spectroscopic Telescope Array (\nustar) consists of two identical co-aligned detectors (FPMA and FPMB) operating over the energy range of 3-79 keV \citep{Ha13}. \nustar~ observed \src\ on 7 August 2024 for an exposure of $\sim$90 ks. Observation details are listed in Table~\ref{tab:log_table_burst}. \nustar~ data are analyzed using the standard pipeline {\tt NUSTARDAS} within {\tt HEASoft} v6.36, together with the latest version of calibration database ({\tt CALDB}) version 20250428. Cleaned event files are produced using {\tt NUPIPELINE}. Source events are extracted from a circular region of 80 arcsec radius centered on the source. The background events are selected from a source-free region. Light curves for both detectors are generated using {\tt NUPRODUCTS}. The data are barycenter corrected using {\tt barycorr} with the JPL DE405 Solar System ephemeris. Background-subtracted light curves are obtained with {\tt lcmath}.
\section{Results}
\label{res}
The NS LMXB \src~ remains relatively poorly explored. Coordinated observations with \xmm{} and \nustar{} in 2024 revealed a total of 15 thermonuclear bursts, including four events detected simultaneously by both instruments. One burst in the \xmm{} dataset (B3) displays an unusual doublet structure, with a separation of $\sim$20~min between the two peaks. The peak count rate of the \xmm{} bursts lies in the range of 220--270~counts~s$^{-1}$, whereas, for the secondary burst of the doublet, the peak count rate reaches $\sim$170~counts~s$^{-1}$. In the \nustar{} light curve, the burst peak count rate is typically in the range of 140--170 counts~s$^{-1}$, except for the secondary peak of the doublet, where the peak count rate is $\sim$90~counts~s$^{-1}$. It is to be noted that the primary burst of the doublet is not covered in the \nustar~ data. The left panel of Fig.~\ref{fig:burst_combined} presents the 1~s binned light curve obtained from \xmm\ and \nustar\ data, showing 15 bursts from \src, four of which are observed simultaneously by both instruments. The right panel highlights that the \xmm\ B3 burst exhibits a rare doublet profile, a unique and uncommon feature.
\subsubsection{Timing analysis of X-ray bursts}
Since the spin period of \src{} is unknown and no burst oscillations had been reported, we perform a blind search for burst oscillations in individual bursts to constrain the spin of the neutron star. The detailed timing analysis is performed using barycenter-corrected clean event files from the \xmm~ and \nustar~ observations. We employed both Fast Fourier Transform (FFT)-based techniques with Leahy normalization and a $Z_1^2$ search, covering the 100–900~Hz range. The dynamic power density spectrum is computed over the burst interval using a sliding-window search, in which successive overlapping time windows are used to track the temporal evolution of oscillation signals as a function of frequency.
\begin{figure*}
    \adjustbox{valign=t}
{\includegraphics[width=1.1\columnwidth]{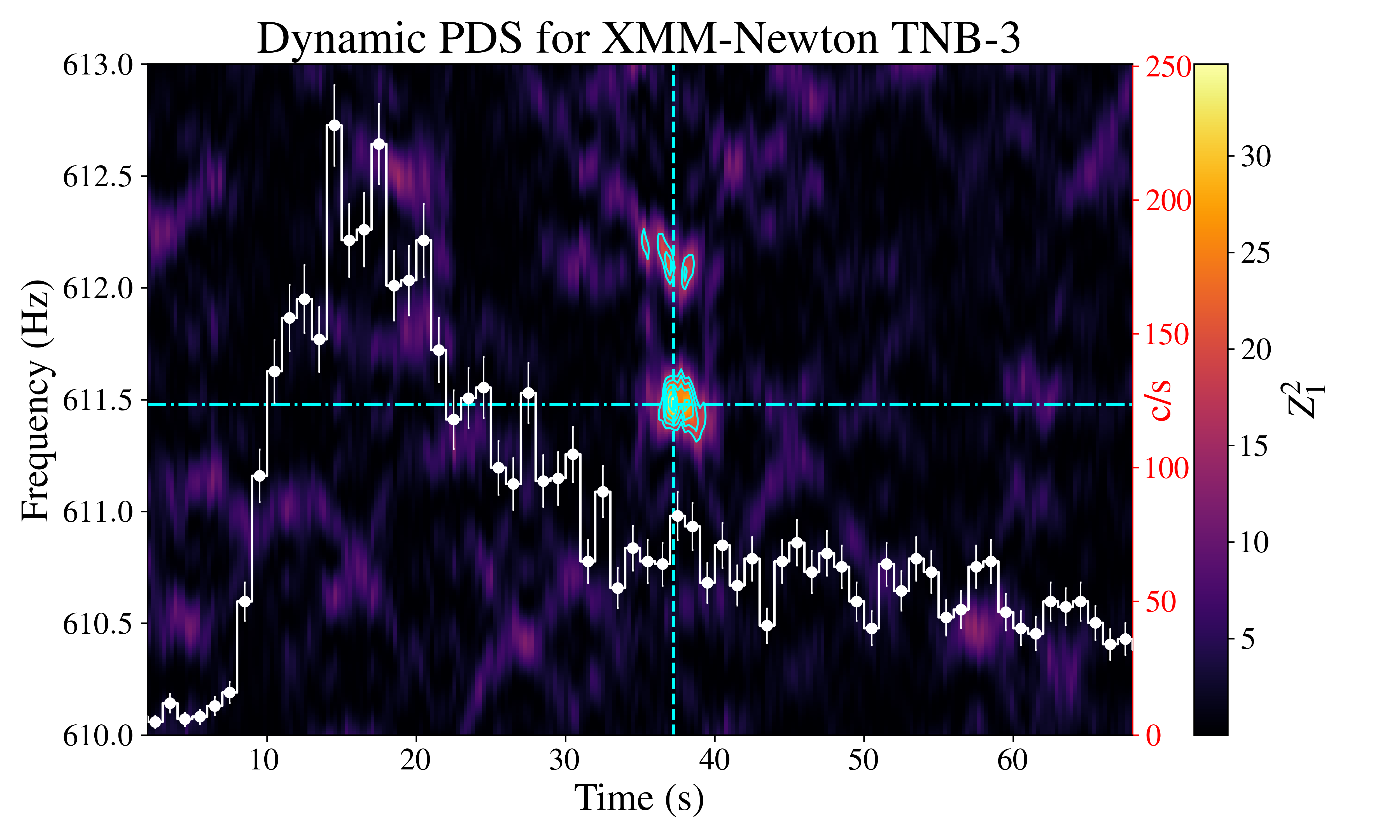}}    
 \adjustbox{valign=t} {\includegraphics[width=\columnwidth]{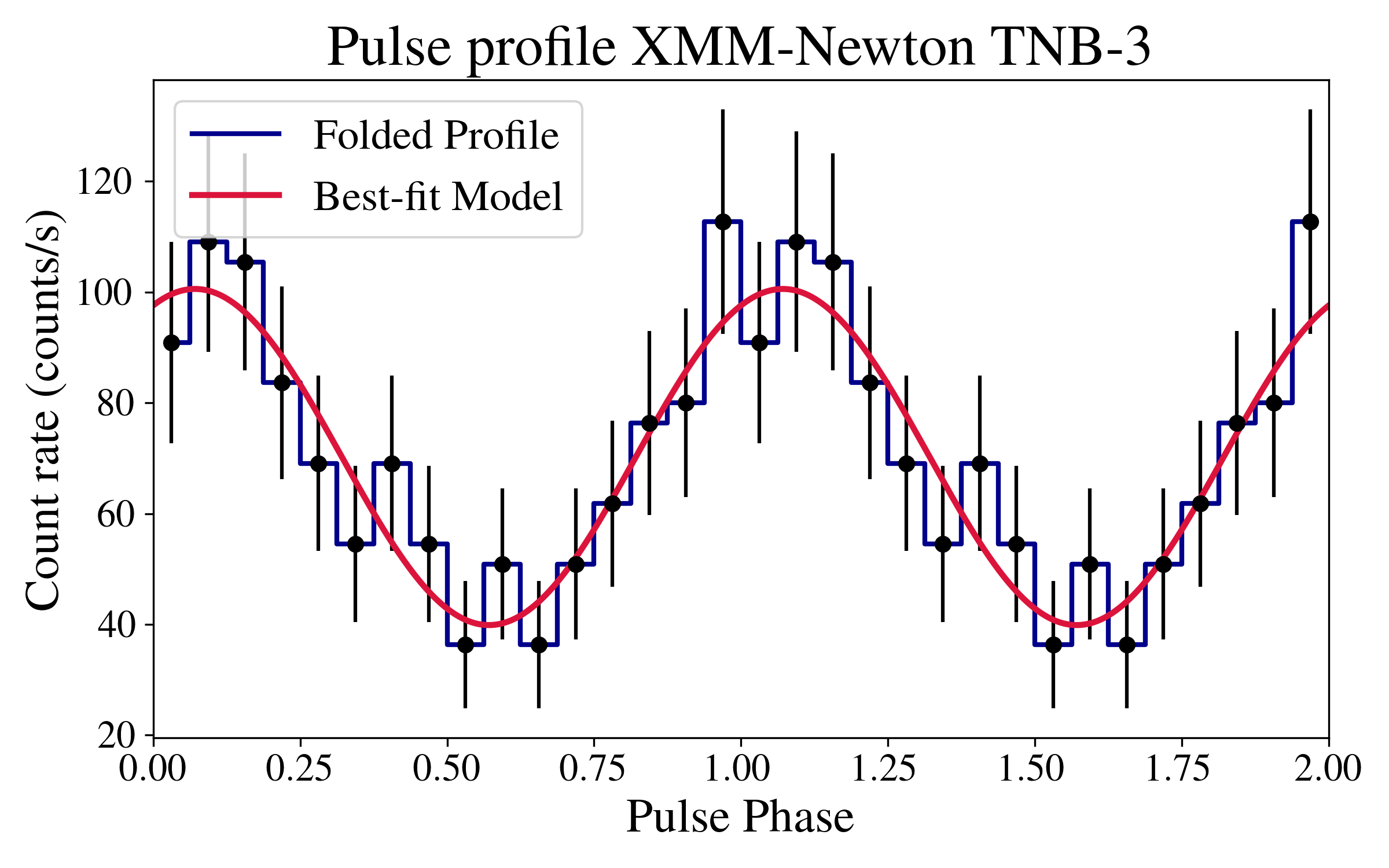}}
\caption{{\bf Left panel:} Dynamic power density spectrum (DPDS) of \src, computed from \xmm\ EPIC-pn data using a 4-s sliding $Z_1^2$ window with a 0.25-s step. The X-axis shows the time evolution of the burst, while the primary Y-axis represents frequency in Hz. The color scale encodes the $Z_1^2$ power. The 1-s binned burst light curve is overlaid in white and corresponds to the secondary Y-axis, which shows the count rate (counts s$^{-1}$). Contours correspond to $Z_1^2$ = 16–36 in steps of 4. The horizontal and vertical dashed cyan lines indicate the spin frequency and the time of maximum power, respectively. {\bf Right panel:} Folded pulse profile of \src{} from a 4-s segment at the peak frequency of 611.48 Hz, along with the best-fitting sinusoidal model.}
    \label{fig:DPDS_XMM}
\end{figure*}
\begin{figure*}
 \adjustbox{valign=t} {\includegraphics[width=\columnwidth]{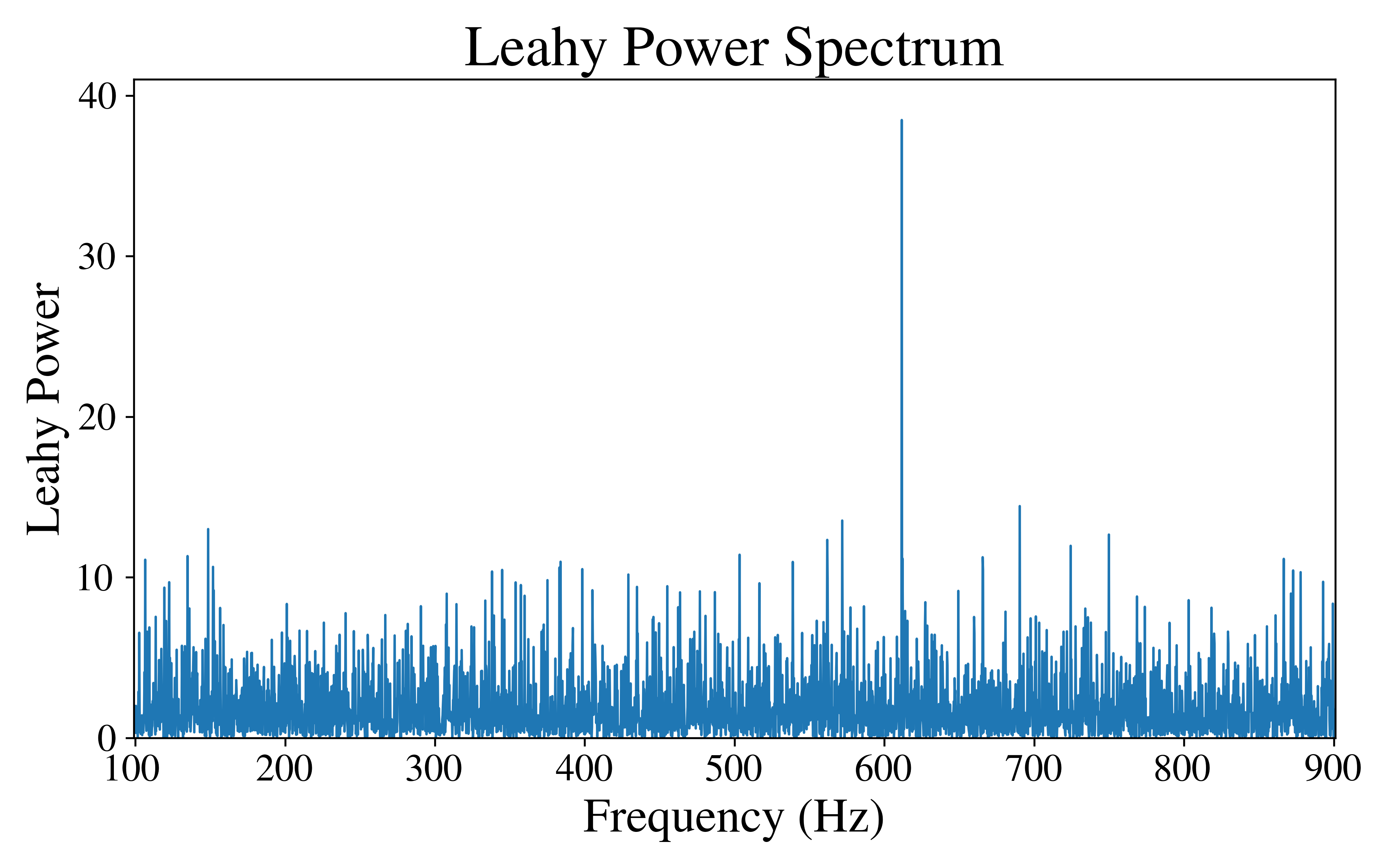}}
   \adjustbox{valign=t}{\includegraphics[width=\columnwidth]{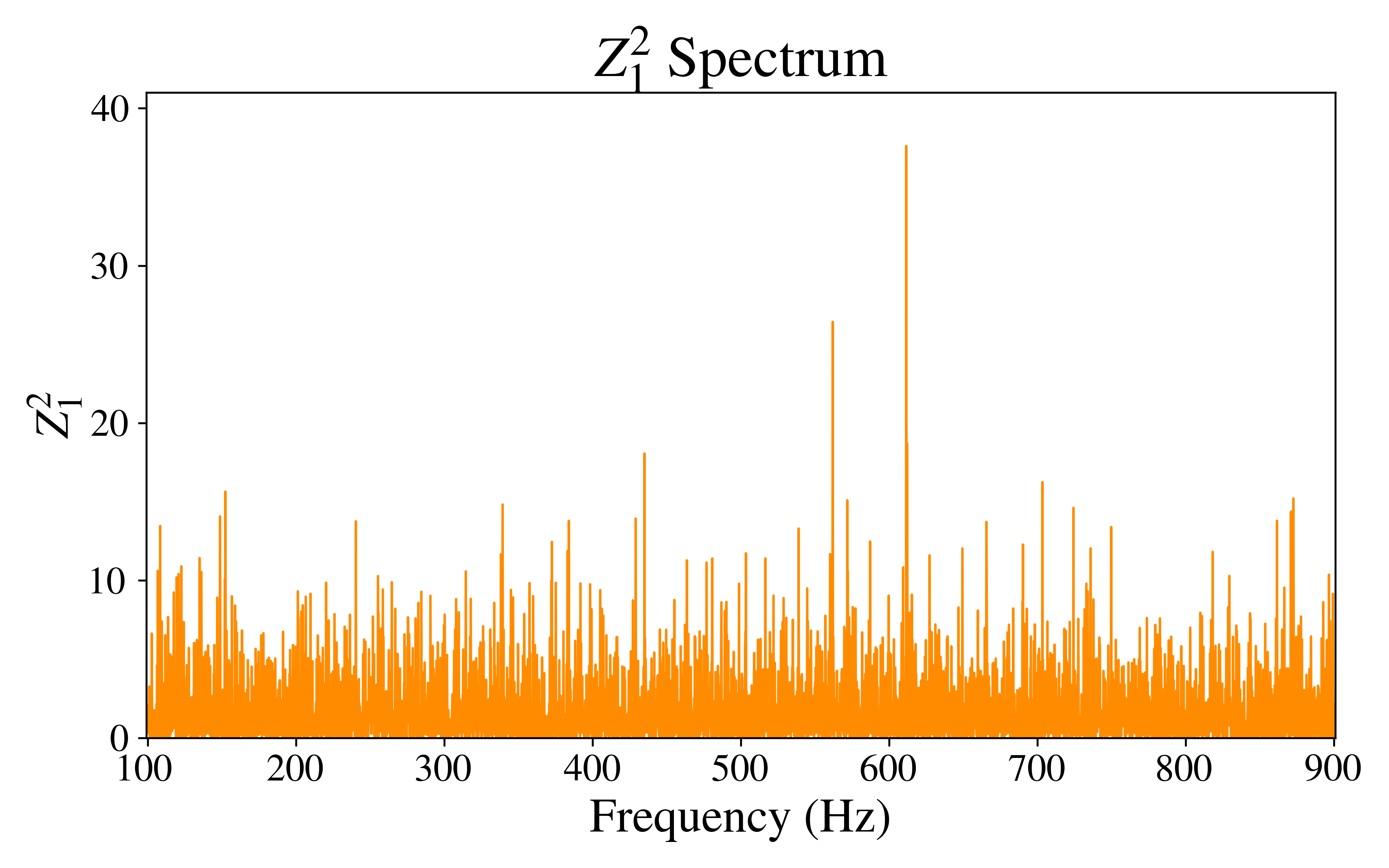}}
\caption{Power spectra from the FFT (left) and $Z_1^2$ (right) analyses showing a signal at $\sim 611.5$~Hz in \xmm~B3. The power spectra are obtained from a 4~s segment around the peak power over 100-900 Hz.}
    \label{fig:PDS_XMM}
\end{figure*}
\subsection{Burst oscillations search}
In this work, we carried out a comprehensive timing analysis of all thermonuclear bursts observed from NS LMXB \src{} using the barycentre-corrected clean event data obtained with \xmm~ and \nustar. In total, 15 bursts are identified across the datasets: 9 with \xmm{} EPIC-pn, and 6 with \nustar. This high-quality dataset motivates us to search for burst oscillations, which have not been reported previously for this source. We analyzed all available burst data from both \xmm{} and \nustar{} in our burst oscillations search. A tentative burst oscillation is detected only during the primary burst of the doublet B3, observed with \xmm. The primary burst of the same doublet, however, was not covered with the \nustar{} observation. For the \xmm{} data, we analyzed barycentre-corrected events in the 0.5--10~keV energy range, while for \nustar, we used the 3--40~keV band. The dynamic PDS search is performed over 100--900~Hz using a 4\,s sliding window with a step size of 0.25~s and frequency resolution of 0.25 Hz, across a burst duration of 70\,s. This analysis reveals a burst oscillation at $\sim$611.5~Hz, with a maximum power of 34.4. After accounting for the number of trials, the probability is $p_{\rm multi} \approx 1.9 \times 10^{-3}$, corresponding to a trial-corrected significance of $\sim$2.9$\sigma$, assuming 3200 independent frequency bins, 17 independent time segments, and a total of 54400 trials.

The overlapping sliding windows introduce correlations, so the assumption of fully independent trials likely overestimates the trial factor. In addition, we repeated the search using a slightly different setup, adopting a 3~s sliding window with a step size of 0.25~s over 100--900~Hz. This configuration independently recovers the same signal at 611.5~Hz in the same time segments, with a maximum $Z_1^2$ power of $\sim$35. After accounting for the number of trials, the probability increases to $p_{\rm multi} \approx 1.5 \times 10^{-3}$, corresponding to a trial-corrected significance of $\sim$3.0$\sigma$, assuming 2400 independent frequency bins and a total of 55200 trials. The search comprises approximately 2400 independent frequency bins across the 100--900~Hz range. The uncertainty in the measured power is estimated assuming that the $Z_1^2$ statistic follows a $\chi^2$ distribution with two degrees of freedom. This uncertainty propagates into the inferred significance. We note that the trial correction assumes all frequency bins and time windows are independent. However, because of the use of overlapping sliding windows, this likely overestimates the number of independent trials. The results of the burst oscillation search in the 100–900 Hz range, obtained using a sliding-window dynamic power spectrum, are summarized in Table~\ref{tab:burst_oscillation}.

Burst oscillations are investigated by computing the $Z_n^2$ statistic \citep{Buccheri1983} directly from photon arrival times. For a trial frequency $\nu$ and $N$ detected photons, the statistic is given by
\begin{equation}
Z_n^2 = \frac{2}{N} \sum_{j=1}^{n}
\left[
\left( \sum_{k=1}^{N} \cos(2\pi j \nu t_k) \right)^2
+
\left( \sum_{k=1}^{N} \sin(2\pi j \nu t_k) \right)^2
\right].
\end{equation}

We focus on the first two harmonics, $Z_1^2$ and $Z_2^2$, to search for oscillatory signals. We also compute the $Z_2^2$ statistic to search for possible harmonic content. However, it did not yield a significant increase in the detection statistic compared to $Z_1^2$, indicating that higher harmonics are negligible and that the pulse profile is adequately described by a single sinusoidal component. In the absence of a true signal, $Z_n^2$ follows a $\chi^2$ distribution with $2n$ degrees of freedom. For the fundamental ($n=1$), the corresponding single-trial false-alarm probability is
\begin{equation}
p = \exp\!\left(-\frac{Z_1^2}{2}\right).
\end{equation}

To correct for multiple testing across time and frequency bins, we estimated the multi-trial probability as
\begin{equation}
p_{\rm multi} = 1 - (1 - p)^{N_{\rm trials}},
\end{equation}
where $N_{\rm trials}$ represents either the total number of evaluated $(t,f)$ bins or the number of effectively independent trials set by the window duration.

In the \xmm{} data, B3 exhibits a doublet burst profile (see Fig.~\ref{fig:burst_combined}). Using \xmm{} EPIC-pn data, we detect a burst oscillation signal during the decay phase of the primary burst of the doublet in B3. The strongest signal is found at $\sim611.5$~Hz in both the $Z_1^2$ and $Z_2^2$ statistics. The centroid frequency is consistent between the two methods, further supporting the robustness of the signal. For the $Z_1^2$ search, the maximum power is $Z_1^2 = 35$ at $\sim611.5$~Hz, corresponding to a trial-corrected significance of ($\approx 3.0\sigma$). The left panel of Fig.~\ref{fig:DPDS_XMM} shows the dynamic power density spectrum of \src{}, computed from \xmm{} EPIC-pn data using a 4~s sliding $Z^2$ window with a 0.25~s step. For clarity, the figure displays the dynamic PDS of \src~ for a 3~Hz range centered at 611.5~Hz to highlight the burst oscillation signal, while the significance is evaluated over the 100--900~Hz range. The burst oscillation properties obtained over a range of 100--900~Hz using a sliding-window dynamic PDS are summarized in Table~\ref{tab:burst_oscillation}. From a comprehensive analysis of thermonuclear X-ray bursts from several neutron star LMXBs, \citet{Bilous2019} reported detection of two candidate signals in \src~ at 415 and 656~Hz with powers of 35.9 and 32.2, respectively. These values are comparable to the maximum $Z_1^2$ value of 34.75 observed for the 611.5~Hz candidate in the present case, although the frequencies are different.

The power density spectra (PDS) are also computed for a 4~s segment of the burst corresponding to peak power. We employed both FFT-based techniques with Leahy normalization and a $Z_1^2$ search to probe the 100--900~Hz range. Both the FFT (Leahy-normalized) and $Z_1^2$ analyses reveal a prominent peak at $\sim$611.5~Hz (611.48--611.50~Hz), with consistent peak powers in the range of 37.6--38.5 in both methods. The power spectra for a 4~s segment corresponding to the maximum power in both tests are shown in Fig.~\ref{fig:PDS_XMM}. For the $Z_1^2$ statistic, the trial-corrected significance is $\sim$3.4$\sigma$ after accounting for $\sim5.4 \times 10^{4}$ trials from the combined frequency and time-domain search. The number of frequency trials ($\sim$3200) is estimated from the 100--900~Hz range with an independent resolution of $1/T \approx 0.25$~Hz for a 4~s segment, while $\sim$17 independent time segments are obtained over the $\sim$70~s burst duration. Similarly, the Leahy-normalized FFT yields a trial-corrected significance of $\sim$3.4$\sigma$ after applying the same trial correction. The consistency between the FFT and $Z_1^2$ analyses, both in frequency and statistical significance, provides evidence for a burst oscillation at $\sim$611.5~Hz.

To further quantify the strength of the detected burst oscillation signal, we compute the fractional rms amplitude, $A_{\rm rms}$. The fractional rms amplitude, $A_{\rm rms}$, is defined as
\begin{equation}
A_{\rm rms} = \sqrt{\frac{P_s}{N_m} \left( 1 - \frac{N_{\rm bkg}}{N_m} \right)},
\end{equation}
where $P_s$ is the signal power, $N_m$ is the total number of detected photons in a given segment, and $N_{\rm bkg}$ represents the background counts. For thermonuclear bursts with high count rates, the background contribution is negligible ($N_{\rm bkg} \ll N_m$), and the expression of the rms amplitude reduces to
\begin{equation}
A_{\rm rms} \simeq \sqrt{\frac{P_s}{N_m}}.
\label{eqn2}
\end{equation}
The signal power is derived from the measured Fourier power, $P_m$, using the noise-subtraction method of \citet{Groth1975}. During the \xmm{} B3 burst, a peak $Z_1^2$ power of $P_m \approx 35$ is detected at $\sim$611.5 Hz. The corresponding interval contains $N_m = 317$ photons, yielding a noise-corrected signal power of $P_s = 30 \pm 11$. Using Equation~\ref{eqn2}, we obtain a fractional rms amplitude of $A_{\rm rms} = (30.8 \pm 5.7)\%$.
 
To further examine the pulse morphology associated with this oscillation, we constructed the folded pulse profile at the detected frequency. The pulse profile obtained by folding the burst data at this frequency is shown in the right panel of Figure \ref{fig:DPDS_XMM}, along with the best-fitting sinusoidal model. The folded pulse profile is fitted with a sinusoidal model $F(t) = A + B\,\sin(2\pi \nu t - \phi_0)$, where $A$ represents the constant (DC) level, $B$ is the modulation amplitude, $\nu$ is the oscillation frequency, and $\phi_0$ is the phase offset. The best-fitting parameters are $A = 70.0 \pm 4.0$, $B = 30.4 \pm 5.6$, and $\phi_0 = -1.1 \pm 0.2$~rad [$(-0.35 \pm 0.06)\,\pi$]. From these values, the fractional rms amplitude of the pulse is derived as $f_{\rm rms} = B/(\sqrt{2}\,A) = 30.5 \pm 5.9\%$, consistent with the amplitude inferred from the $Z^2$ power. The corresponding best-fit model to the folded pulse profile is shown in the right panel of Figure~\ref{fig:DPDS_XMM}. The relatively high rms amplitude during the decay phase is atypical and may indicate residual asymmetry in the emission, though limited photon statistics could partially drive it.  
\begin{figure}
\includegraphics[width=\columnwidth] {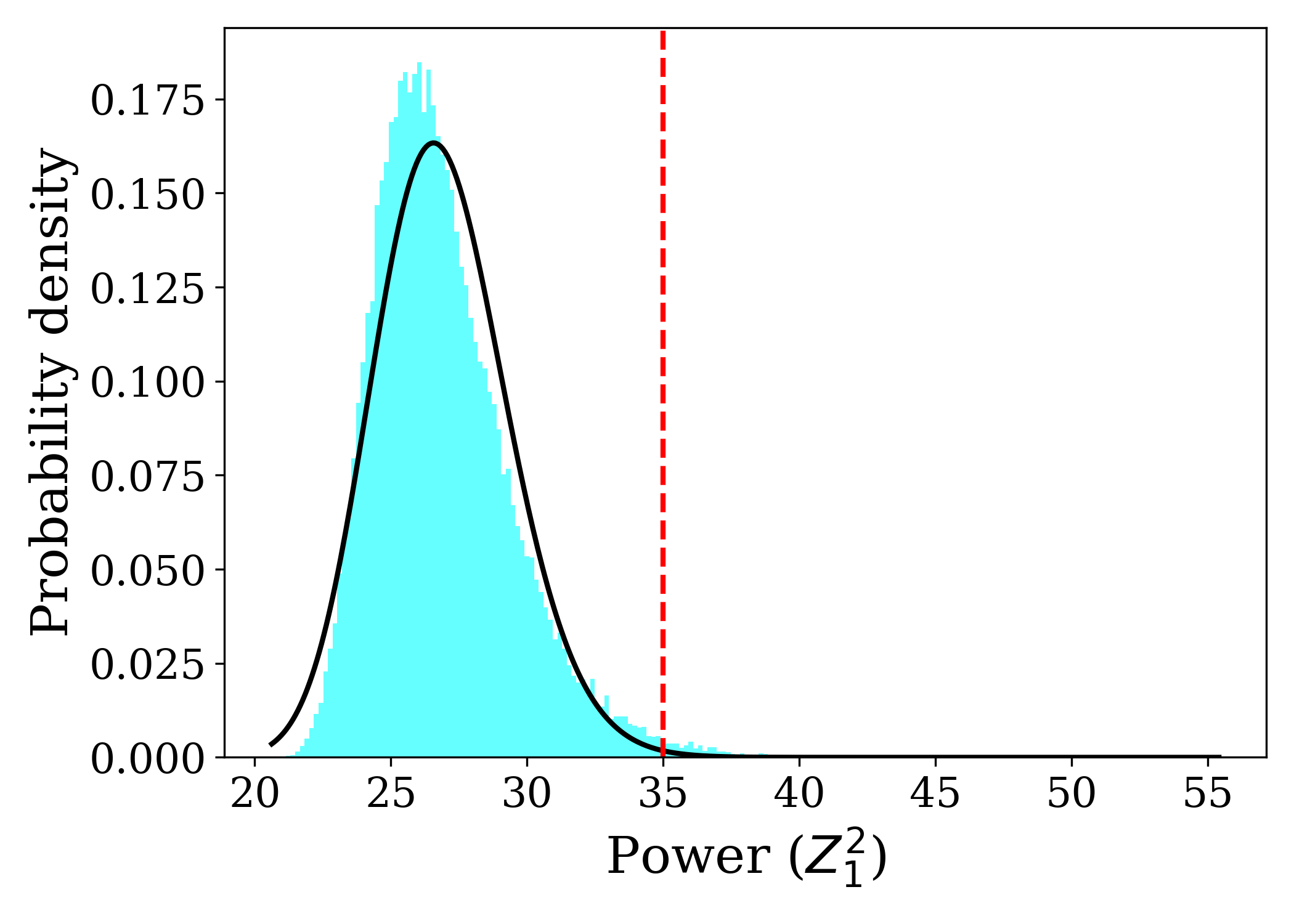}
\caption{ Distribution of the maximum $Z_1^2$ values obtained from 50000 Monte Carlo simulations of the burst for a 100-900 Hz search. The black curve shows the posterior log-normal fit to the simulated maxima, while the red vertical line indicates the observed maximum $Z_1^2$ measured from the \textit{XMM-Newton}/EPIC-pn burst (B3).}
    \label{fig:MCMC_XMM}
\end{figure}
\begin{table*}
\centering
\caption{Burst oscillation properties of \src{} are investigated using dynamic power density spectra obtained from the \xmm/EPIC-pn data. A signal at $\sim$611.5~Hz is consistently recovered using different window configurations and in an independent blind search over 100--900~Hz.}
\begin{tabular}{lcccccc}
\hline
Burst no. & $\nu$ (Hz) & $Z_1^2$ power & Window size (s) & Step size (s) &$p_\textrm{ind}$ & Significance  \\ 
\hline
B3 &   611.50 & 34.4 & 4 & 0.25 & $1.9\times10^{-3}$  & 2.9$\sigma$  \\
B3 &   611.50 & 34.9 & 3 & 0.25 & $1.5\times10^{-3}$  & 3.0$\sigma$ \\
\hline
\label{tab:burst_oscillation} 
	\end{tabular}
\end{table*}
\subsubsection{Simulations of X-ray burst light curves}
The statistical significance of the burst oscillation signal is evaluated using Monte Carlo (MC) simulations of the burst light curves. Since timing analysis employs overlapping time windows that are not statistically independent, analytical estimates of the false-alarm probability are not reliable. To properly account for correlations introduced by overlapping segments and the non-stationary nature of the burst emission, we determined the significance numerically. The same timing-search procedure is applied to the observed and simulated data, and for each realization, we record the maximum $Z_1^2$ power. The fraction of simulations that yield a maximum power at or above the observed value provides an estimate of the false-alarm probability, which is then converted to equivalent Gaussian significance.

Following established approaches \citep{Bilous2019, Li2022, Bult2021, Mandal2026ApJ, Zanon2025}, we generate synthetic burst light curves by randomly redistributing photon arrival times according to the observed count-rate profile. For each simulated light curve, we apply the same burst-oscillation search procedure described in Section~3.1. We perform a search over 100–900 Hz using a 3~s sliding window with a step size of 0.25~s and a frequency resolution of 0.25 Hz. This method preserves the burst morphology and Poisson noise properties, thus providing a realistic representation of the null hypothesis. Each simulated light curve is processed identically to the real data. For the primary burst of the doublet in the \xmm{} burst B3, we perform $5\times10^{4}$ simulations to construct the distribution of maximum $Z_1^2$ values. The simulated peak appears at $Z_1^2 \sim 26.9$ for the 100-900 Hz search, whereas the observed value ($P_m = 35$) lies in the extreme tail of the simulated distribution, indicating that the signal is unlikely to arise by chance.

When the search is extended over the 100--900 Hz range, we obtain $p_{\rm emp}$ of \(8.9 \times 10^{-3}\), corresponding to a significance of \(\sim 2.37\sigma\). The posterior median significance in this case is $\sim 2.92\sigma$. Figure~\ref{fig:MCMC_XMM} shows the simulated distribution together with its best-fitting log-normal approximation for the 100--900 Hz search based on 50,000 Monte Carlo simulations. Since the simulations explicitly account for the adopted frequency range and sliding-window search procedure, the derived probabilities inherently include the associated trial factors. Although global significance decreases for the wider frequency search, it is unlikely that the detected feature is produced solely by random statistical fluctuations. 
\section{Discussion and Conclusions}
\label{dis}
The neutron star low-mass X-ray binary \src{} is a relatively poorly studied source. Joint observations with \xmm{} and \nustar{} in 2024 detected a total of 15 thermonuclear bursts, of which four are observed simultaneously with both instruments. One burst observed with \xmm{} exhibits a rare doublet burst profile, a phenomenon previously reported in only a few sources (e.g., EXO~0748-67, 4U~1636-536, 4U~1608-522). The separation between the primary and secondary peaks of the doublet burst profile in the \xmm{} observation (B3) is $\sim$20~min. Thermonuclear bursts occurring on very short timescales pose a challenge to standard ignition models, as short recurrence times are insufficient for significant fresh fuel accumulation. These events typically show a strong initial burst followed by a weaker secondary, although cases with similar fluence or even multiple closely spaced bursts have also been reported. It remains uncertain whether such behavior shares a common origin with double-peaked bursts or represents a separate class of phenomena. A likely interpretation is that the secondary burst is fueled by a fraction of material left unburnt during the primary burst, which may later ignite after being transported to deeper layers. This extensive, high-quality burst sample makes \src{} an excellent candidate for investigating burst oscillations, which had not previously been detected in this source. We present the results of a detailed timing analysis of all thermonuclear bursts from \src{}.

We identify a possible burst oscillation signal from \src\ at $\sim$611.5 Hz, reaching a maximum $Z_{1}^{2}$ power of $\sim$35 during the \xmm{} burst-3. The signal has a trial-corrected global significance of $\sim$3.0$\sigma$. To assess the robustness of the signal, we performed Monte Carlo simulations, which yielded an empirical global significance of $\sim 2.4\sigma$. In the \xmm{} observation, the oscillations are detected during a burst exhibiting a rare doublet profile, with significant and consistent power in both the $Z_{1}^{2}$ and $Z_{2}^{2}$ statistics, supporting the presence of a coherent signal. The oscillation strength, pulse-profile morphology, and fractional rms amplitudes are broadly consistent with those observed in other accreting millisecond X-ray pulsars and neutron star LMXBs \citep{Ga08}. However, the fractional rms amplitude measured during the decay phase is relatively high compared to typical values. This may indicate a residual asymmetry in the emission pattern, although limited photon statistics could partially influence it. Independent estimates of the fractional rms amplitude, derived from both pulse-profile modeling and the $Z^{2}$ statistic, are in good agreement, further supporting the robustness of the detection. The oscillation frequency corresponds to a neutron-star spin period of $\sim$1.6 ms ($\sim$611.5 Hz), suggesting that \src\ may harbor a rapidly rotating neutron star comparable to those in 4U~1608$-$522 ($\sim$620 Hz; \citet{Ga08}) and 4U~1820$-$30 ($\sim$720 Hz; \citet{Ja24}). 

Doublet bursts are generally attributed to incomplete fuel burning followed by rapid re-ignition. The resulting non-uniform burning and residual fuel distribution can produce surface asymmetries that may enhance the detectability of oscillations, although burst oscillations are not uniquely associated with doublet events. Sources exhibiting complex burst morphologies, such as doublet or double-peaked profiles (e.g., 4U~1636-536 \citep{Roy2022} and 4U~1705-44 \citep{Giri2026}), have also been observed to show burst oscillations. These multi-peaked structures are commonly interpreted as signatures of non-uniform flame spreading or temporary stalling of the burning front on the neutron star surface. While the exact relationship between burst morphology and oscillation mechanisms remains unclear, such behavior may reflect conditions conducive to the generation of burst oscillations.

The physical origin of the brightness asymmetries responsible for burst oscillations is not yet fully established. These asymmetries are generally thought to arise from localized regions on the neutron-star surface where thermonuclear burning ignites and spreads. However, the detailed geometry and propagation of such hot spots remain uncertain. Despite this, burst oscillations provide valuable insight into the burning processes and the composition and structure of the neutron star’s outer layers \citep{Strohmayer2006}. In several neutron star LMXBs (e.g., 4U~1728-34, 4U~1636-536, KS~1731-260, X~1658-298), the oscillation frequency is observed to drift by a few Hz during the course of a burst \citep{Watts2012, Muno2002, Bostanci2023}, with the largest reported drift reaching $\sim$5~Hz in X~1658-298 \citep{Wijnands2001}. This behavior is commonly attributed to angular-momentum conservation in an expanding and contracting burning layer, or to variations in the pattern speed of the emitting region \citep{Watts2012, Strohmayer1997}. In contrast, no significant frequency drift is observed in the burst oscillations from \src, indicating a relatively stable oscillation. More broadly, burst oscillations can appear at different phases of a thermonuclear burst and are likely associated with distinct physical processes, including ignition asymmetries and flame spreading during the rise \citep{Strohmayer1997}, rotational modulation of surface emission \citep{Strohmayer2006, Watts2012}, and the persistence of asymmetries driven by hydrodynamic instabilities or global ocean modes during the decay \citep{Spitkovsky2002, Cumming2000}. Variations in the accretion rate can influence the initial conditions for thermonuclear ignition and, consequently, the detectability of burst oscillations. 

However, it remains unclear why some neutron stars in AMXPs exhibit burst oscillations while others do not, and even in sources where they are present, they are not observed in every burst. In several neutron star low-mass X-ray binaries, burst oscillations are highly intermittent, appearing only in a small fraction of detected bursts. For instance, oscillations are observed in only one out of fourteen bursts from 4U~1916-053 \citep{Ga08}. Similarly, in 4U~1636-536, burst oscillations were detected in just three out of fifteen bursts, with one of these exhibiting a doublet burst profile \citep{Roy2022}. Another potential source is 4U~1730-22, where a strong signal at 584.65 Hz was detected for the first time in only a single \nicer~ burst out of sixteen observed during 2021-2022 \citep{Li2022}. This highlights that burst oscillations are highly transient and unpredictable, likely depending on several underlying physical conditions. The large oscillation amplitudes observed during the early phase of some bursts are broadly consistent with a spreading hot-spot scenario \citep{Strohmayer1997}. In contrast, the presence of oscillations in the burst tail is more difficult to explain, as the burning layer is expected to have spread over the entire neutron star surface by that stage. These late-time oscillations are therefore often attributed to brightness asymmetries generated by hydrodynamic instabilities \citep{Spitkovsky2002} or to surface modes excited in the neutron star ocean \citep{Cumming2000}.
\facilities{ADS, HEASARC, \xmm{}, \nustar{}}

\software{HEASoft V6.36 \citep{heasoft}, XSPEC V12.13.0 \citep{Ar96}}, NumPy and SciPy \citep{virtanen20}, Matplotlib \citep{hunter07}, IPython \citep{perez07}. 
\section*{Acknowledgements}
We thank the referee for his/her valuable comments and suggestions, which improved the manuscript. The authors sincerely thank Francesco Coti Zelati and Tolga Güver for their valuable comments and insightful suggestions. The research work at the Physical Research Laboratory, Ahmedabad, is funded by the Department of Space, Government of India. This research has made use of data obtained with \nustar{}, a project led by Caltech, funded by NASA, and managed by NASA/JPL, and has utilized the {\tt NUSTARDAS} software package, jointly developed by the ASDC (Italy) and Caltech (USA). We acknowledge the use of public data from the \xmm{}, and \nustar{} data archives. 
\section*{Data Availability}
The data used for this article are publicly available in the High Energy Astrophysics Science Archive Research Centre (HEASARC).

\bibliography{4U1323}{}
\bibliographystyle{aasjournal}

\end{document}